\documentclass[a4paper, onecolumn, prX, showkeys]{revtex4}
\usepackage{graphicx}
\usepackage{amsmath, amsfonts, amsthm, mathrsfs, enumitem}

\newcommand{\norm}[1]{\left|\left|#1\right|\right|}
\newcommand{\seminorm}[1]{\left|#1\right|}
\newcommand{\abs}[1]{\left|#1\right|}

\newtheorem*{def1.1}{Definition 1.1}
\newtheorem*{con1.2}{Conjecture 1.2}
\newtheorem*{def2.1}{Definition 2.1}
\newtheorem*{lem2.2}{Lemma 2.2}
\newtheorem*{lem2.3}{Lemma 2.3}
\newtheorem*{lem2.4}{Lemma 2.4}
\newtheorem*{lem2.5}{Lemma 2.5}
\newtheorem*{thm2.6}{Theorem 2.6}

\def\R{{\mathbb R}}

\def\T{{\mathcal{T}}}
\def\eps{{\varepsilon}}

\def\u{{u}}
\def\f{{f}}
\def\l{{\ell}}
\def\d{{\,\mathrm{d}}}
\def\paper{{Anand20051789}}
\def\qiao{{PhysRevLett.106.085504}}

\begin{document}

\title{Resolution of a Conjecture in Nonlocal Strain-gradient Plasticity}

\author{Jordan S. Cotler}
\affiliation{Massachusetts Institute of Technology, Cambridge, Massachusetts 02139, USA}
\author{Felipe Hern{\'a}ndez}
\affiliation{Massachusetts Institute of Technology, Cambridge, Massachusetts 02139, USA}

\begin{abstract}
Strain-gradient theories of plasticity have been successful in modeling 
the behavior of complex materials.  
However, the traditional formulation
of these theories lacks a material length scale, and is thus incapable of 
capturing experimentally observed size effects that play an important role
in the behavior of nano structures.  As a result, a modified theory was proposed which incorporates an intrinsic dissipative
length scale.  The theory predicts that the solutions to the flow rule are global minimizers of the functional for energy dissipation.  We prove that there are no global minimizers of the functional, thus resolving a previously unsolved conjecture.  Our result shows that the variational formulation of the theory is unviable.  The non-existence of a global minimizer appears to be 
related to the formation of infinitely fine plastic boundary layers.
\end{abstract}

\keywords{Strain-gradient Plasticity, Calculus of Variations}

\maketitle

\section{Introduction}
In traditional formulations of strain-gradient plasticity, stresses and strains
are decomposed into elastic and plastic components.
For the one-dimensional small strain model considered in this paper, the
decomposition is of the form
$$
\gamma = \gamma^e + \gamma^p,
$$
where $\gamma$ denotes the strain, $\gamma^e$ denotes the elastic component,
and $\gamma^p$ denotes the plastic component.  In one dimension, the 
elastic strain is spatially constant and is directly proportional to the 
applied stress $\tau$.  The plastic deformation is given by solving the flow
rule
$$
\left.
\begin{aligned}
    \tau = S\left(\frac{d^p}{d_0}\right)^m \frac{\dot{\gamma}^p}{d^p}, 
         &\quad \,\,\,d^p = |\dot{\gamma}^p|, \\
    \dot{S} = H(S)d^p,\,\,\,\,\,
            &\quad S(y,0) = S_0,
\end{aligned}
\right\}
$$
where
\begin{enumerate}[label=(\roman{*})]
    \item $d^p$ is an effective flow rate
    \item $d_0$ is a reference flow rate
    \item $m$ is a rate sensitivity parameter
    \item $S$ is an internal state variable that represents resistance to 
        plastic flow
    \item $H(S)$ is a hardening function.
\end{enumerate}
Despite the effectiveness of strain-gradient theories, they are incapable of describing
size-dependent effects due to the lack of an inherent length scale.
In an attempt to fix this deficiency, Anand et. al. proposed the use of a
dissipative length scale $\l$ \cite{\paper}.  The length scale $\l$ is 
incorporated naturally into the effective flow rate $d^p$ by 
$$
d^p = \sqrt{\seminorm{\dot{\gamma}^p}^2 + \l^2 \seminorm{\dot{\gamma}_{,y}^p}^2},
$$
where subscripts denote spatial derivatives.
With this approach, the problem of finding the yield stress for a sample of
length $h$ is given by the solution to the differential equation
\begin{equation}
    \label{diffeq}
    \frac{\tau}{S_0} = \frac{\u}{\sqrt{|\u|^2 + l^2|\u_{,y}|^2}}
    - \l^2 \frac{\partial}{\partial y}\left[\frac{\u_{,y}}{\sqrt{|\u|^2 + \l^2|\u_{,y}|^2}}\right]
\end{equation}
subject to the constraints
\begin{equation}
    \label{constraints}
    \u(0) = \u(h) = 0, \quad \frac{1}{h}\int_0^h u \d y  = 1,
\end{equation}
where $u=\dot{\gamma}$ is the flow rate and $y$ denotes the spatial variable.
For convenience, we give the following definition:
\begin{def1.1}
    A function $u$ is \emph{admissible} if it satisfies \eqref{constraints}.
\end{def1.1}

The purpose of this paper is to
show that in the case $\ell/h > \sqrt{2}-1/2$, solutions
to Equation~\eqref{diffeq} cannot be minimizers to the associated energy
\begin{equation}
    \label{energy}
    \mathcal{T}(\f) := \frac{S_0}{h}\int_0^h\sqrt{\f(y)^2 + \l^2 |\f(y)_{,y}|^2} \d y,
\end{equation}
thus disproving the conjecture stated in \cite{\paper}.  Note that the Euler-Lagrange equation for the functional in (3) is given by (1).
The conjecture is as follows:
\begin{con1.2}
    \label{conj}
    The energy~\eqref{energy} has a minimum value over the space
    of admissible functions.  Moreover, this minimum value corresponds to the 
    yield strength $\tau$, and, in addition, any minimizing function
    satisfies Equation~\eqref{diffeq}.
\end{con1.2}

\section{Proofs of main results}
The idea of the proof is that if $\ell/h>\sqrt{2}-1/2$, and 
$u$ is an admissible function satisfying
the Euler-Lagrange equation, then we can construct a new admissible function
$v^*$ that has a lower energy than $u$.  Therefore, $u$ cannot minimize the
energy $\T$.

\begin{def2.1}
    Let us define the functional $\mathcal{I}$ by
    $$
    \mathcal{I}(\f) := \frac{S_0\,\l}{h} \left(|\f(0)| + |\f(h)|\right) + \T(\f).
    $$
\end{def2.1}
\begin{lem2.2}
Let $u$ be an admissible function, and let $v$ be a function on $[0,h]$ with 
$\mathcal{I}(v) < \mathcal{T}(u)$.  Then there exists an admissible function 
$v^*$ such that $\mathcal{T}(v^*) < \mathcal{T}(u)$.
\end{lem2.2}
\begin{proof}
In what follows, $D$ denotes a weak spatial derivative.
Let 
$$
v_\eps(x)  = 
	\left\{
     \begin{array}{ll}
       \frac{u(0)}{\eps} y,  &y \in [0, \eps]\\ \\
       v(\frac{h(y-\eps)}{h-2\eps}),  &y \in [\eps, h-\eps] \\  \\
       \frac{u(h)}{\eps} (h-y),  &y \in [h-\eps, h]
     \end{array}
   \right.
$$ 
We will set $\tilde{v}_\eps = K_\eps v_\eps$, 
where $K_\eps$ is the normalization constant 
$$
K_\eps := \frac{h}{h-2\eps+\eps\cdot(v(0)+v(h))/2}
$$
chosen so that $\tilde{v}_\eps$ is admissible.
Now we show that $\lim_{\eps\to0}\mathcal{T}(\tilde{v}_\eps) = \mathcal{I}(v)$.
We have
\begin{align}
    \label{nastyint}
\mathcal{T}(\tilde{v}_\eps)
= \frac{S_0}{h}K_\eps\left[\int_0^\eps \sqrt{v_\eps^2 + \l^2Dv_\eps^2}\d y + 
\int_{h-\eps}^h\sqrt{v_\eps^2 + \l^2Dv_\eps^2}\d y
+ \int_{\eps}^{h-\eps} \sqrt{v_\eps^2 + \l^2Dv_\eps^2}\d y \right].
\end{align}
Because we have $\lim_{\eps\to 0}K_\eps = 1$, let us only deal with the terms inside 
the brackets.  The first integral in \eqref{nastyint} can be simplified to 
$$
\int_0^\eps \sqrt{v_\eps^2 + \l^2Dv_\eps^2} \d y
= \frac{\abs{v(0)}}{\eps} \int_0^\eps \sqrt{\l^2 + y^2} \d y.
$$
Since $l \leq \sqrt{l^2+y^2} \leq \l+\eps$ for all $y < \eps$, we have upon
integration that 
$$
\lim_{\eps\to 0} \frac{\abs{v(0)}}{\eps} \int_0^\eps \sqrt{\l^2 + y^2} \d y 
= \l\abs{v(0)}.
$$
Similarly, the third integral in \eqref{nastyint} tends to $\l\abs{v(h)}$. 
Finally, we consider the remaining integral 
$$
\frac{S_0}{h}\int_{\eps}^{h-\eps} \sqrt{v_\eps^2 + Dv_\eps^2} \d y
= \frac{S_0}{h-2\eps} \int_0^h \sqrt{v^2 + 
\left( \frac{h}{h-2\eps}\, \l \, Dv \right)^2} \d y.
$$
The integrand is bounded below by $\sqrt{v^2+\l^2(Dv)^2}$ and above by 
$\frac{h}{h-2\eps}\sqrt{v^2+\l^2(Dv)^2}$, so in the limit $\eps\to 0$,
$$
\frac{S_0}{h-2\eps} \int_0^h \sqrt{v^2 + 
\left( \frac{h}{h-2\eps} \,\l \, Dv \right)^2} \d y \to \mathcal{T}(v).
$$
Therefore, $\mathcal{T}(v_\eps) \to \mathcal{I}(v)$.  Since 
$\mathcal{I}(v) < \mathcal{T}(u)$, we can find an $\eps$ small enough
that, setting $v^*=v_\eps$, we have $\mathcal{T}(v^*) < \mathcal{T}(v)$
as desired.
\end{proof}

The function $v^*$ constructed in the previous lemma is not necessarily twice differentiable.
However, this is not a problem because by the inequality 
$$
\sqrt{a^2+b^2} \leq \abs{a} + \abs{b}
$$
we obtain 
$$
\abs{\mathcal{T}(v) - \mathcal{T}(u)} < \l\norm{u-v}_{1,1}.
$$
That is, the functional $\T$ is continuous in the $W^{1,1}$ norm.
Thus, by the density of 
$C^\infty$ functions in $W^{1,1}$, we can find a smooth function $v_\infty^*$ 
approximating $v^*$ such that $\mathcal{T}(v_\infty^*) < \mathcal{T}(u)$.

We now establish an a priori bound for $\T(\f)$ over the space of 
admissible functions.
\begin{lem2.3} 
    \label{apriori}
    For any admissible function $\f$,
    $\T(\f) > \frac{S_0}{\sqrt{2}} (1+2\frac{\l}{h})$.
\end{lem2.3}
\begin{proof}
We first bound the infimum of the related functional
$$
\mathcal{N}(\f) := \frac{S_0}{h}\int_0^h \abs{\f} + \l \abs{D\f} \d y
$$
over the set of admissible functions.  Since $\f$ is admissible,
$$
\frac{S_0}{h}\int_0^h \abs{\f} \d y \geq S_0.
$$
Now, consider the integral $\int_0^h \l\abs{D\f} \d y$, which is the total variation 
of $\f$, for which we obtain the bound
$$
\int_0^h \l\abs{D\f} \d y \geq 2\l \max_{y\in[0,h]} \f(y) > 2\l.
$$
The second inequality is obtained by the consideration that 
if $\max_{y\in[0,h]} \f(y) \leq 1$, then $\f$ is not admissible.
Combining the above bounds, we obtain
$\mathcal{N}(\f) \geq S_0 (1 + 2\frac{\l}{h})$.  
Since
$$
\T(\f) = \int_0^h \sqrt{\f^2 + \l^2(D\f)^2} \d y \geq 
\int_0^h \frac{1}{\sqrt{2}}\left(|\f|+\l|D\f|\right) \d y
= \frac{1}{\sqrt{2}}\,\mathcal{N}(\f),
$$
it follows that
$\mathcal{T}(\f) > \frac{S_0}{\sqrt{2}} (1+2\frac{\l}{h})$.
\end{proof}

\begin{lem2.4}
Let $\ell/h > \sqrt{2} - 1/2$, and let $u$ be an admissible 
function such that there exists an $m\in\R^+$ with 
$u(y) < my$ on some $(0, \delta) \subset [0, h]$, and $u(\delta) = m\delta$.  
Then the function
$$
v(y)  = 
	\left\{
     \begin{array}{ll}
       K m \delta,  &y < \delta\\ \\
       K u(y),  &y \geq \delta
     \end{array}
   \right.
$$ 
has
$\mathcal{I}(v) < \mathcal{T}(u)$ where $K$ is a normalization constant
chosen such that $\frac{1}{h}\int_0^h v(y) \d y = 1$. 
\end{lem2.4}
\begin{proof}
Let $C = m\delta$ for convenience.
We can write $K = \frac{h}{h+C\delta - A}$, 
where we define
$$
A := \frac{1}{h}\int_0^\delta u \d y.
$$
We have 
$$
\T(u) = \frac{S_0}{h}\int_0^h \sqrt{u^2 + \l^2(Du)^2}\d y
$$
and
\begin{align*}
    \mathcal{I}(v) &= \frac{S_0\,\l}{h} |v(0)| + \frac{S_0}{h}\int_0^\delta KC  \d y
    + \frac{S_0}{h}\int_\delta^h K \sqrt{v^2 + \l^2(Dv)^2} \d y\\
    &= K\frac{S_0}{h}\left(\l C + C \delta + \int_\delta^h \sqrt{v^2 + \l^2(Dv)^2} \d y\right).
\end{align*}
We need $\mathcal{I}(v) < \T(\u)$, so we want to show that $\T(\u) - \mathcal{I}(v) > 0$.
Using the above expressions, we get the following equation:
\begin{align*}
\frac{\mathcal{T}(u) - \mathcal{I}(v)}{K}
&= \frac{1}{h}\mathcal{T}(u)(C\delta - A)  - \frac{S_0}{h}C\delta + 
\frac{S_0}{h}\left(\int_0^\delta \sqrt{u^2 + \l^2(Du)^2} \d y - \l C\right). 
\end{align*}
The term in parentheses is positive because 
$$
\int_0^\delta \sqrt{u^2 + \l^2(Du)^2} \d y 
\geq \int_0^\delta \l \abs{Du} \d y \geq \l u(\delta) = \l C.
$$
We now turn our attention to the term 
$\frac{1}{h}\T(\u)(C\delta - A) - \frac{S_0}{h}C\delta$.  Since $\u < my$ on
$(0,\delta)$, we have the bound
$$
A = \frac{1}{h}\int_0^\delta u\d y < \frac{1}{2} m\delta^2  =\frac{1}{2} C \delta.
$$
Applying this to the term of interest yields
$$
\frac{1}{h}\mathcal{T}(u)(C\delta-A) - \frac{S_0}{h}C\delta 
\geq \frac{1}{2h}\mathcal{T}(u) \, C\delta - \frac{S_0}{h}C\delta
= \frac{1}{h}C\delta \left(\frac{1}{2} \mathcal{T}(u) - S_0\right).
$$
Let us now show that the term in parentheses in the above equation is positive.  
We can use the bound from
the previous lemma to obtain
$$
\frac{1}{2} \mathcal{T}(u) - S_0
\geq \frac{S_0}{2\sqrt{2}}\left(1+2\frac{\l}{h}\right) - S_0.
$$
Since by hypothesis $1+2 \ell/h > 2\sqrt{2}$, it follows that
$\frac{1}{h}\T(\u)(C\delta - A) - \frac{S_0}{h}C\delta$ is positive.
Therefore, we can conclude that
$$
\frac{\mathcal{T}(u) - \mathcal{I}(v)}{K} > 0
$$
and seeing that $K>0$, the above equation gives us $\T(\u) > \mathcal{I}(v)$.
\end{proof}

\begin{lem2.5}
    \label{absurdity}
    Let $u$ be an admissible function that satisfies the Euler-Lagrange
    equation \eqref{diffeq}.  Then $u_{,y}(0) = 0$.  
\end{lem2.5}
\begin{proof}
The Euler-Lagrange equation gives us
$$
\frac{\lambda}{h} + 
\frac{S_0}{h} \frac{u + \l^2 u_{,yy}}
{\sqrt{u^2 + \l^2 (u_{,y})^2}} - 
\frac{S_0}{h} 
\frac{\l^2 \left(u_{,y} \right)^2
\left(u + \l^2 u_{,yy} \right)}
{\left(u^2 + \l^2 (u_{,y})^2 \right)^{\frac{3}{2}}}=0.
$$
Since the energy $\T$ is independent of $y$, we can also use the first integral of 
the Euler-Lagrange equation and get that
$$
\frac{S_0}{h} \sqrt{u^2 + \l^2(u_{,y})^2} + \frac{\lambda}{h} u - \frac{S_0}{h} 
\frac{\l^2 (u_{,y})^2}{\sqrt{u^2 + \l^2(u_{,y})^2}}=B
$$
where $B$ is some real constant.  Let us assume by contradiction that $u_{,y}(0) 
\not = 0$.  At $u(0) = 0$, the above two equations give $\lambda=0$ and $B=0$.  
Plugging these constants into either of the above two equations gives $u=0$ for 
all $y$ which is not admissible.  Therefore, we have a 
contradiction, and so the only remaining possibility is that $u_{,y}(0)=0$.
\end{proof}

\begin{thm2.6}
If $\ell/h > \sqrt{2} - 1/2$, then no admissible minimizer to \eqref{energy}
solves the Euler-Lagrange equation
\eqref{diffeq}.
\end{thm2.6}
\begin{proof}
Suppose that $u$ is an admissible function that minimizes $\mathcal{T}(u)$ and 
satisfies the Euler-Lagrange equation.  Then by the previous lemma,
we can conclude that $u_{,y}(0) = 0$.  
We will now use the previous results to demonstrate that there exists an
admissible function $v^*$ such that $\mathcal{T}(v^*) < \mathcal{T}(u)$, and 
thereby reach a contradiction to Conjecture 1.2.

In order to construct $v^*$, we show that $u$ satisfies the conditions of 
Lemma 2.4.
Consider the line $f(y)=\frac{1}{2}y$.  First, note
that for sufficiently small $\eps>0$, $u(\eps) < \frac{\eps}{2}$
because $(u - f)_{,y} = -\frac{1}{2} < 0$.  
Furthermore,
$u$ must intersect the line $f$, otherwise it cannot satisfy 
$\frac{1}{h}\int_0^h u \d y = 1$.  
Choosing the smallest $\delta$ such that $u(\delta) = f(\delta)$, 
we find that $u$ satisfies the conditions of Lemma 2.4, as desired. 
Therefore, there exists an admissible function $v^*$
such that $\mathcal{T}(v^*) < \mathcal{T}(u)$.  It follows that $u$ cannot be a 
minimizer of $\mathcal{T}$, and so we have a contradiction.
\end{proof}

We remark that the nonlocal extension to power law plasticity was developed for the case in which $\ell/h$ tends to zero, which corresponds to microscale behavior.  Thus, the technical constraint $\ell/h > \sqrt{2} - 1/2$ is sufficient to demonstrate the failure of the variational principle in the context of the theory.  

\section{Conclusion}
We have shown that the variational formulation of the nonlocal extension to power-law plasticity theory
proposed by Anand et. al. in \cite{\paper} is not mathematically viable 
because the energy dissipation functional assigns a finite energy to 
discontinuous functions.  Thus, minimizers are not smooth and do not satisfy 
the Euler-Lagrange equation.  The discontinuities in the minimizer appear to be 
manifestations of infinitely fine plastic boundary layers.
However, we note that the modification to the theory proposed
by Qiao et. al. in \cite{\qiao} is well-posed.
Briefly, this modification introduces an energetic length scale, adding 
a term $\l_e (u_{,y})^2$ to the integrand of the energy $\T$.  This new energy is
convex in $u_{,y}$ and 
coercive in the $H^1$ norm.  A standard argument in the direct method for 
the calculus of variations verifies that minimizers exist. 

\section*{Acknowledgements}

The authors wish to thank Gigliola Staffilani for her feedback and useful discussions.

\end{document}